\documentclass[letterpaper,11pt]{emulateapj}
\usepackage{epsfig}
\usepackage{natbib}
\usepackage{amsmath}

\begin{document}
\newcommand{\kms}{km~s$^{-1}$}
\newcommand{\Msun}{M_{\odot}}
\newcommand{\Lsun}{L_{\odot}}
\newcommand{\ML}{M_{\odot}/L_{\odot}}
\newcommand{\etal}{{et al.}\ }
\newcommand{\hhh}{h_{100}}
\newcommand{\hsq}{h_{100}^{-2}}
\newcommand{\tn}{\tablenotemark}
\newcommand{\mdot}{\dot{M}}
\newcommand{\p}{^\prime}
\newcommand{\kmsMpc}{km~s$^{-1}$~Mpc$^{-1}$}

\title{Two Planes of Satellites in the Centaurus A Group}

\author{R. Brent Tully$^{1}$}
\author{Noam I. Libeskind$^{2}$}
\author{Igor D. Karachentsev$^{3}$}
\author{Valentina E. Karachentseva$^{4}$}
\author{Luca Rizzi$^{5}$}
\author{Edward J. Shaya$^{6}$}

\affil{$^1$Institute for Astronomy, University of Hawaii, 2680 Woodlawn Drive, Honolulu, HI 96822, USA}
\affil{$^2$Leibniz-Institut f\"ur Astrophysik Potsdam, An der Sternwarte 16, D-14482, Potsdam}
\affil{$^3$Special Astrophysical Observatory, Nizhnij Arkhyz, Karachai-Cherkessian Republic, 369167 Russia}
\affil{$^4$Main Astronomical Observatory, National Academy of Sciences of Ukraine, Kiev, 03680 Ukraine}
\affil{$^5$W. M. Keck Observatory, 65-1120 Mamalahoa Hwy., Kamuela, HI 96743, USA}
\affil{$^6$Astronomy Dept., University of Maryland, College Park, MD 20743, USA}

\begin{abstract}
Tip of the red giant branch measurements based on Hubble Space Telescope and ground-based imaging have resulted in accurate distances to 29 galaxies in the nearby Centaurus~A Group.
All but two of the 29 galaxies lie in either of two thin planes roughly parallel with the supergalactic equator.  The planes are only slightly tilted from the line-of-sight, leaving little ambiguity regarding the morphology of the structure.   The planes have characteristic r.m.s. long axis dimensions of $\sim 300$ kpc and short axis dimensions of $\sim 60$~kpc, hence axial ratios $\sim 0.2$, and are separated in the short axis direction by 303 kpc.

\end{abstract}

\section{Introduction}

There is convincing evidence that half of the Messier~31 satellites lie in a thin plane \citep{2013ApJ...766..120C, 2013Natur.493...62I}.  It was argued by \citet{2013MNRAS.436.2096S} that most of the rest of the M31 satellites lie in a second, almost parallel, plane.  These latter authors derived plausible orbits for the galaxies in the two planes.  It was suggested that the satellites were born in a strata above M31 within the emerging Local Void and collectively chased M31 into the Local Sheet with the expansion of the void.  The Milky Way has played a shepherding role.

Preliminary evidence has been presented \citep{2015AJ....149...54T} that something similar may be happening in the nearby Centaurus A (Cen A) Group.  The group is rich in satellites and almost all of them appear to lie within either of two thin planes.  This letter presents the observational evidence for this claim in greater detail.

The Cen A Group is very nearby, with the peculiar elliptical Cen A itself at 3.66 Mpc.  \citet{1998A&AS..127..409K, 2000A&AS..146..359K} have carried out intense surveys for possible group members and follow up observations with Hubble Space Telescope (HST) have resulted in accurate distance measurements for many targets from the luminosities of resolved stars at the tip of the red giant branch (TRGB).  Properties of the group have been discussed by \citet{2002A&A...385...21K} and \citet{2005AJ....129..178K}.

The accurate distances obtained with HST using the TRGB method have clarified a problem created by a projection effect.  A group around Messier 83 partially overlaps on the sky with the Cen A group and completely overlaps in velocities, but distance measurements reveal that the M83 Group is to the background with a clearly resolved 1~Mpc gap.  Maps of the projected and three-dimensional distributions of galaxies in the adjacent groups are to be found in \citet{2015AJ....149...54T}.

The effort to obtain accurate distances and velocities of galaxies in the Cen~A region is ongoing.  A small number of known galaxies still exist that potentially may be group members.  Observations in HST cycle 21 program 13442 have lead to four new distances. Two galaxies are confirmed as members of the Cen~A Group: KKs53 and KK203 at 2.93 and 3.78 Mpc respectively.  One galaxy, NGC 5264, is a member of the M83 Group at 4.78 Mpc.  The fourth galaxy ESO 270-17 is also known as Fourcade-Figueroa \citep{1971BAAA...16...10F} and is found only $3^{\circ}$ from Cen~A.   This large and low surface brightness galaxy is of interest, but it lies in the background with no known neighbors at 6.94 Mpc.  The reduction procedures that lead to these distance measurements are most recently described by Karachentsev et al. (2015). Color-magnitude diagrams and TRGB fits for the newly observed targets and for all the previous cases are made available in the CMDs/TRGB catalog at the Extragalactic Distance Database.\footnote{http://edd.ifa.hawaii.edu}  

Recently, distances for two dwarfs in close proximity to Cen~A from an ongoing survey with the Magellan Clay 6.5 m telescope have been reported by  \citet{2014ApJ...795L..35C}. 
Table~1 identifies all galaxies confirmed with distance measurements to lie in the Cen~A group and known possible members that lack distance confirmations.

\begin{table}
\scriptsize{
\caption{Members and Posibble Members of Cen A Group}
\begin{tabular}{r|c|c|c|r|c|c|c}
\hline
PGC No. & Name & D   & Vhel & Ty & $M_B$ & SGL & SGB \\
        &      & Mpc & km/s &    & mag   & deg & deg \\
\hline
Plane 1 \\
\hline
  46957 &  NGC5128    & 3.66 & 547 & -2 & -20.54 & 159.7529 & -5.2494\\
  45279 &  NGC4945    & 3.72 & 560 &  6 & -19.33 & 165.1793 &-10.2161\\
  46674 &  NGC5102    & 3.74 & 467 & -1 & -17.84 & 153.4179 & -4.0673\\
  47171 &  ESO324-024 & 3.78 & 525 & 10 & -15.47 & 158.3895 & -4.4130\\
  45717 &  ESO269-058 & 3.75 & 400 & 10 & -15.14 & 162.9226 & -8.8312\\
  45917 &  NGC5011C   & 3.73 & 647 & -2 & -14.01 & 159.3867 & -7.4704\\
  45916 &  ESO269-066 & 3.75 & 784 & -1 & -13.86 & 160.9727 & -7.8860\\
  46680 &  KK197      & 3.84 &     & -3 & -12.29 & 159.1091 & -5.7166\\
  46663 &  KK196      & 3.96 & 741 & 10 & -12.04 & 161.5396 & -6.4539\\
  45104 &  ESO269-037 & 3.15 & 744 & -3 & -11.62 & 162.2485 & -9.9058\\
2815820 &  KKs53      & 2.93 &     & -3 & -10.42 & 155.0355 & -6.7223\\
 166158 &  KK189      & 4.23 &     & -3 & -10.86 & 157.9678 & -7.1880\\
 166167 &  KK203      & 3.78 &     & -3 & -10.33 & 162.1021 & -5.5762\\
2815822 &  KKs055     & 3.85 &     & -3 & -10.05 & 159.3040 & -5.7388\\
        &  MM-Dw1     & 3.63 &     & -3 & -10.1  & 158.9330 & -4.0769\\
        &  MM-Dw2     & 3.60 &     & -3 &  -7.6  & 158.8977 & -4.1211\\
\hline
Plane 1?  \\
\hline
  45628 &  PGC45628   &      & 681 & 10 & -13.06 & 143.5511 & -3.9565\\
  44110 &  ESO219-010 & 4.8: &     & -4 & -12.62 & 165.5514 &-11.8156\\
 166164 &  KK198      &      &     & -3 & -10.47 & 150.5421 & -2.9953\\
2815821 &  KKs54      &      &     & -3 & -10.24 & 148.8408 & -2.7877\\
\hline
Plane 2  \\
\hline
  48334 &  NGC5253    & 3.55 & 403 &  8 & -17.19 & 149.8146 &  1.0062\\
  49050 &  ESO383-087 & 3.19 & 326 &  8 & -16.55 & 154.6356 &  1.3283\\
  47762 &  NGC5206    & 3.21 & 577 & -3 & -16.12 & 165.1068 & -5.3769\\
  48139 &  NGC5237    & 3.33 & 361 & -3 & -14.82 & 160.2663 & -3.0706\\
  48738 &  ESO325-011 & 3.40 & 543 & 10 & -14.02 & 159.7855 & -1.4605\\
  48515 &  KK211      & 3.68 & 600 & -5 & -11.99 & 162.7605 & -3.0833\\
4689187 &  CenN       & 3.66 &     & -3 & -10.93 & 165.3378 & -2.8987\\
 166175 &  KK217      & 3.50 &     & -3 & -10.67 & 163.4596 & -2.5533\\
 166179 &  KK221      & 3.82 & 507 & -3 & -10.52 & 164.8401 & -2.6023\\
2815823 &  KKs057     & 3.83 &     & -3 & -10.21 & 160.2545 & -2.2933\\
 166172 &  KK213      & 3.77 &     & -3 &  -9.80 & 161.4948 & -2.3517\\
\hline
Plane 2?  \\
\hline
  48937 &  KKs59      &      & 688 & 10 & -15.77 & 170.7716 & -4.9166\\
2815824 &  KKs58      &      &     & -4 & -10.76 & 154.6544 &  0.6160\\
\hline
Other  \\
\hline
  39032 &  ESO321-014 & 3.28 & 611 & 10 & -12.63 & 152.2641 &-17.6207\\
  51659 &  P51659     & 3.62 & 390 & 10 & -11.85 & 166.9467 &  3.8360\\
\hline
Other?  \\
\hline
2815819 &  KKs51      &      &     &  0 & -11.50 & 157.9323 &-12.5448\\
\hline
\end{tabular}
\label{table1}
}
\end{table}

\section{Two Planes}

\begin{figure}[]
\begin{center}
\includegraphics[scale=0.75]{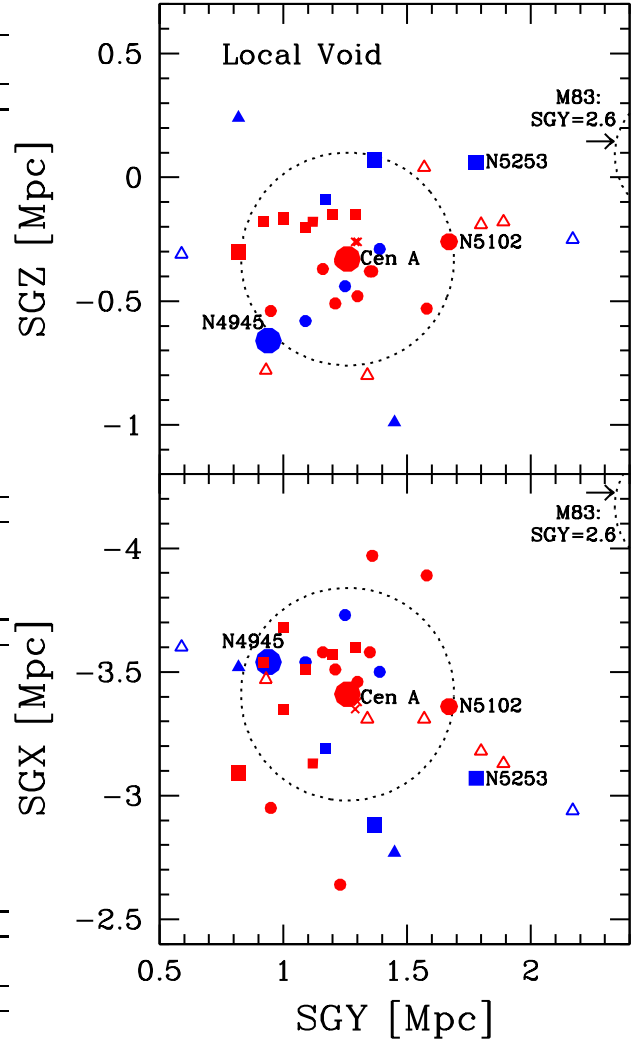}
\caption{Two projections of galaxies in the Centaurus A Group. Galaxies with accurate (unknown) distances are given filled (open) symbols; red for E/S0 and blue for spirals/irregulars. Planes 1, 2, or other galaxies are represented by circles, squares, and triangles, respectively. The projected second turnaround (virial) radius around Cen A is shown by the dotted circle.  The proximity of the M83 Group is indicated by the partial circles at the right boundaries.  An apparent separation into two planes is seen in the top panel.} 
\label{xyz}
\end{center}
\end{figure}

The distribution of galaxies in the Cen~A Group are shown in two supergalactic projections in Figure~\ref{xyz}.  It is to be recalled that the equator of the supergalactic system was defined \citep{1964rcbg.book.....D} by the concentration of galaxies to a band on a scale of 3,000~\kms.  Galaxies within 7~Mpc are strongly confined to the Local Sheet \citep{2008ApJ...676..184T} which is coincident with the supergalactic equator.  Returning to Fig.~\ref{xyz}, the galaxies with known distances are represented by filled symbols, 14 circles, 11 squares, 2 overlapping crosses, and 2 triangles. Typical 5\% TRGB distance errors of 175~kpc at the Cen~A Group project largely in the SGX direction and almost not at all in the SGZ direction.  It can be seen in the top panel that the circles and squares appear to lie in two separate, roughly parallel bands that are slightly slanted from horizontal.  The band of circles will be called plane~1 and the band of squares will be called plane~2.  Only the two galaxies represented by triangles, out of 29 Cen~A group members with good distances, are clearly not associated with one of these two bands.   There are six candidates for Cen~A membership that lack distance measurements. The galaxy ESO219-010 has a surface brightness fluctuation distance of 4.78~Mpc \citep{2000AJ....119..166J}  that places it in the background but the distance is uncertain and this galaxy is also retained as a candidate.  These seven potential members are plotted as open triangles in Fig.~\ref{xyz} assuming the distance of Cen~A.  Four of these are plausible members of plane~1, two could be in plane~2, and only one cannot be assigned tentatively to either of the two planes.  We conclude that the argument of this paper will not be fundamentally changed with new distance measurements.
The two dwarfs from the \citet{2014ApJ...795L..35C} survey represented by the almost merged crosses adjacent Cen~A qualify as members of plane 1 but will not be considered further.  The Crnojevi\'c et al. survey is restricted to an area around Cen~A that only reaches the edge of plane 2 so does not contribute to a fair sample of the full group.

A familiar pattern is seen in the distribution of galaxies around Cen~A.  E/S0/dSph are more centrally concentrated than spirals and irregulars. Every E/S0/dSph is within or at the boundary of the estimated second turnaround circle, an approximation to the virial radius \citep{2015AJ....149...54T}.  Spirals and irregulars are more dispersed, yet they can be assigned to one or other of the planes.  The second major galaxy in the group, the spiral NGC~4945, is located at the edge of the virial domain.  It appears to belong to plane~1.

If the planes have a physical significance, it would be fortuitous if they are being viewed exactly edge on from our vantage point.  A search for an optimal viewing orientation will be carried out next.  However evidently we are near an optimal viewing direction.  As a consequence, an evaluation of the reality of the planes is hardly affected by distance uncertainties because uncertainties project essentially within the planes.  Galaxies lacking distance measures are retained in order to evaluate if the analysis might fundamentally change with their addition.

\section{A Cen~A Reference Frame}

It is evident from inspection that the normals to the two posited planes are similar.  A quantitative evaluation leads to the conclusion that the best fit normal directions differ by only $7^{\circ}$ and that this difference is not statistically significant.   We accept an average of the two normal directions, weighted by the numbers of plane members, and find that it differs from the SGZ axis by $17^{\circ}$.\footnote{Varying distances randomly within 5\% of measured values, the mean normal values differ from the SGZ axis in the range $15.6^{\circ}-20.6^{\circ}$ for 99.7\% of 10,000 trials.}  The direction along the derived normal to the two planes will be called CaZ. The origin of the Cen~A reference frame is taken to coincide with Cen~A. By construction, the CaX and CaY axes are parallel to the planes.   There are no meaningful extensions within the planes, so these two axes are taken in directions that approximate supergalactic coordinates.  CaX is $\sim 4^{\circ}$ from SGX with positive values toward our location,  and CaY is $\sim 16^{\circ}$ from SGY with positive values toward the Virgo Cluster.

The coordinate transformation is achieved with
\begin{equation}
CaX = R_{xx} SGX_c + R_{xy} SGY_c + R_{xz} SGZ_c
\end{equation}
\begin{equation}
CaY = R_{yx} SGX_c + R_{yy} SGY_c + R_{yz} SGZ_c  
\end{equation}
\begin{equation}
CaZ = R_{zx} SGX_c + R_{zy} SGY_c + R_{zz} SGZ_c .
\end{equation}
The coordinate transformation is centered on Cen~A
\begin{equation}
SGX_c = SGX + 3.41
\end{equation}
\begin{equation}
SGY_c = SGY - 1.26
\end{equation}
\begin{equation}
SGZ_c = SGZ + 0.33 .
\end{equation}
The rotation matrix is
\begin{equation}
R = \begin{bmatrix} +0.994 & -0.043 & +0.102\\ -0.001 & +0.919 & +0.393\\ -0.111 & -0.391 & +0.914 \end{bmatrix}
\end{equation}

\begin{figure}[!]
\begin{center}
\includegraphics[scale=0.43]{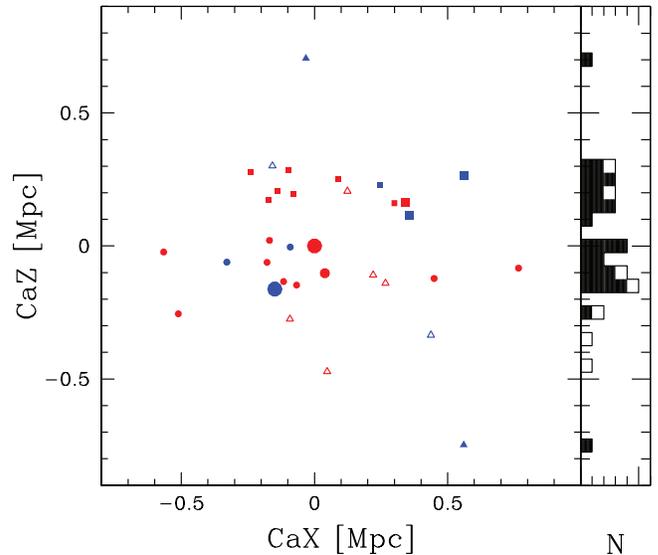}
\caption{Edge-on view of planes.  Symbol shapes and colors are same as in Fig. \ref{xyz}.  Filled histogram: CaZ distribution of galaxies with measured distances.  Open: possible group members without distance measurements.} 
\label{trans}
\end{center}
\end{figure}

An edge-on view after the transformation is shown in Figure~\ref{trans}.    Considering only the galaxies with distance measures and the best fit normals for the two planes considered separately, the r.m.s. dimensions on the longest to shortest axes are 346 to 73 kpc for plane 1 and 250 to 46 kpc for plane 2; minor to major axial ratios of 0.21 and 0.19, respectively.  With the averaged common normal, the r.m.s. plane thicknesses are 77 kpc for plane 1 and 55 kpc for plane 2.  These thicknesses may be inflated by distance errors.  A 5\% error at the Cen~A distance is 175~kpc and the CaZ projection of such an error is $\sim 50$ kpc.

The offset between the means of the two planes is 303 kpc.  The gap in CaZ between the top of plane 1 and the bottom of plane 2 is 94 kpc.

\begin{figure}[!]
\begin{center}
\includegraphics[scale=0.36]{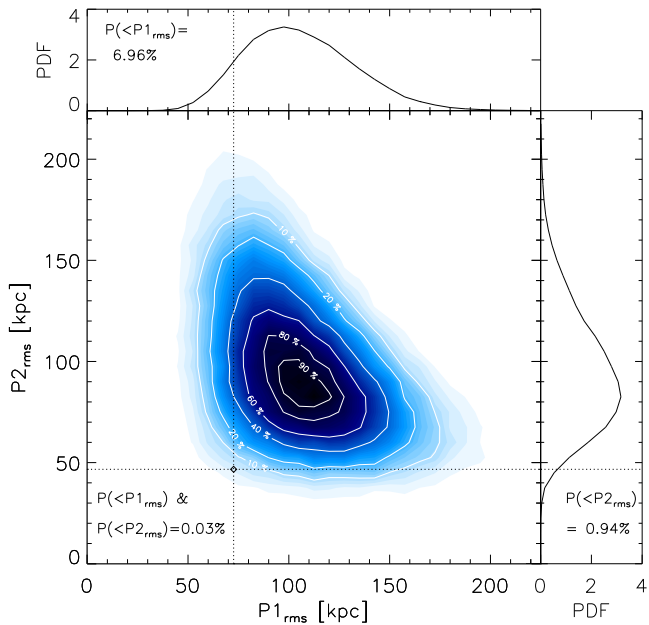}
\caption{
Significance of finding two planes of satellites with values of the minor axis r.m.s. as low as observed. The contours represent the fraction of random Cen A analogues with a given pair of r.m.s. values for two parallel planes, drawn from 100,000 sets that maintain the Cen A radial distribution and global intrinsic flattening. The white lines include labels that indicate the percentage of random draws within these contours. The observed pair of r.m.s. values is shown by the small black diamond, where the two dotted lines intersect. Probability distributions of the r.m.s. values are plotted along the top (plane 1, P1$_{rms}$, probability of 7\%) and at right (plane 2, P2$_{rms}$, probability of 0.9\%). Probability of finding two planes with these low values of the r.m.s. is 0.03\%.}
\label{signif}
\end{center}
\end{figure}

If it is accepted that the distribution of almost all group members is separable into two components, then Figure~\ref{signif} evaluates the likelihood that the observed distributions are compatible with random yet flattened sets of points that maintain the same radial distribution as the satellites of Cen~A. The test involves the construction of 100,000 Cen~A analogs, where the azimuthal and longitudinal positions of Cen~A satellites are randomized, but their original radial positions and the intrinsic flattening of the system as a whole is kept fixed. For each of these 100,000 Cen~A analogues, we separate the satellites into two groups, that have the same number of members as plane 1 and plane 2, but which also minimizes the r.m.s. about two best fit parallel planes. The r.m.s. values about these two planes are plotted against each other as blue contours in Fig.~\ref{signif}. The labels indicate the percentage of random draws that are within each white contour. The probability of randomly finding two planes with r.m.s. values less than or equal to the observed values is 0.03\%.  The probability is even less if the assumption of the global intrinsic flattening is dropped.

\section{Summary}

While modest coordinate transforms peak up the contrast of major to minor axes of the posited planes in the Cen A Group, the splitting into two thin strata is already evident in projection, with distance information serving simply to exclude the confusion caused by non-members.  The two planes lie almost parallel with the supergalactic equator.  If it might seem unlikely that we would be in such a favorable viewing position edge-on to the planes, it is to be recalled that we, too, are in the flattened distribution of galaxies on the supergalactic equator.  Indeed, on scales from the historic Local Supercluster at 40~Mpc to the Local Sheet at 7~Mpc, down now to a scale of 300~kpc, structures are nested with similarly oriented aspects.  The two planes of satellites around M31 \citep{2013Natur.493...62I, 2013MNRAS.436.2096S} are also nearly parallel to the supergalactic equator.  

While the issue of {\it two} planes remains to be explained, there is a probable cause for the organization in the flattened orientations.  The suspected agent lurks in the upper left corner of the top panel of Figure~\ref{xyz}.  All of the galaxies around us within 7 Mpc are part of a wall of the Local Void.  The void is expanding, causing the wall, our Local Sheet, to descend toward negative SGZ at 260~\kms\  \citep{2008ApJ...676..184T} (Karachentsev et al. 2015).  One possible scenario is that initially there were two proto-groups at significantly different distances from the center of the Local Void but headed outward in roughly the same direction.  Both entities would have been stretched out parallel to the SGZ-plane by radial repulsion from the void, and they could either be just now coming together or had a crossing, in which case Cen~A would have strongly influenced the distributions of both planes.  A high fraction of the satellites in plane 2 are dSph suggesting, indeed, that there has been a crossing.  It is worth noting that NGC~4945 is not a negligible player.  This galaxy, in plane 1 at 480~kpc from Cen~A, has a luminosity that is 1/3 that of Cen~A.  Otherwise there is no other immediate influence until M83 and its companions 1~Mpc away (which, by the way, aligns in projection with plane 1: a shepherd?).  The brightest galaxy in plane 2, NGC~5253, is fully 20 times fainter than Cen~A.  If plane 2 were viewed as a separate group, it would have the properties of one of the associations of dwarfs discussed by \citet{2006AJ....132..729T}.

In the case of the planes around M31, \citet{2013MNRAS.436.2096S} reconstructed orbits suggesting that the satellites were formed in strata early in the development of the void and that a memory of the initial conditions has been preserved.  
To have a chance at a similar reconstruction with the Cen~A Group there is a need for a reasonably complete set of velocities.  Currently, as seen in Table~1, many velocities are unknown.  The present discussion is limited to providing evidence that almost all the galaxies in the Cen~A Group lie in two almost parallel thin planes embedded and close to coincident in orientation with planes on larger scales. The two-tiered alignment is unlikely to have arisen by chance.

\bigskip\noindent
{\it Acknowledgment.} Observations and support for this program has been provided by the Space Telescope Science Institute in connection with Hubble Space Telescope program GO-13442.

\bibliography{paper}

\begin{thebibliography}{14}
\expandafter\ifx\csname natexlab\endcsname\relax\def\natexlab#1{#1}\fi

\bibitem[{{Conn} {et~al.}(2013){Conn}, {Lewis}, {Ibata}, {Parker}, {Zucker},
  {McConnachie}, {Martin}, {Valls-Gabaud}, {Tanvir}, {Irwin}, {Ferguson}, \&
  {Chapman}}]{2013ApJ...766..120C}
{Conn}, A.~R., {Lewis}, G.~F., {Ibata}, R.~A., {Parker}, Q.~A., {Zucker},
  D.~B., {McConnachie}, A.~W., {Martin}, N.~F., {Valls-Gabaud}, D., {Tanvir},
  N., {Irwin}, M.~J., {Ferguson}, A.~M.~N., \& {Chapman}, S.~C. 2013, \apj,
  766, 120

\bibitem[{{Crnojevi{\'c}} {et~al.}(2014){Crnojevi{\'c}}, {Sand}, {Caldwell},
  {Guhathakurta}, {McLeod}, {Seth}, {Simon}, {Strader}, \&
  {Toloba}}]{2014ApJ...795L..35C}
{Crnojevi{\'c}}, D., {Sand}, D.~J., {Caldwell}, N., {Guhathakurta}, P.,
  {McLeod}, B., {Seth}, A., {Simon}, J.~D., {Strader}, J., \& {Toloba}, E.
  2014, \apjl, 795, L35

\bibitem[{{de Vaucouleurs} {et~al.}(1964){de Vaucouleurs}, {de Vaucouleurs}, \&
  {Shapley}}]{1964rcbg.book.....D}
{de Vaucouleurs}, G.~H., {de Vaucouleurs}, A., \& {Shapley}, H. 1964,
  {Reference catalogue of bright galaxies}

\bibitem[{{Fourcade}(1971)}]{1971BAAA...16...10F}
{Fourcade}, C.~R. 1971, Boletin de la Asociacion Argentina de Astronomia La
  Plata Argentina, 16, 10

\bibitem[{{Ibata} {et~al.}(2013){Ibata}, {Lewis}, {Conn}, {Irwin},
  {McConnachie}, {Chapman}, {Collins}, {Fardal}, {Ferguson}, {Ibata}, {Mackey},
  {Martin}, {Navarro}, {Rich}, {Valls-Gabaud}, \&
  {Widrow}}]{2013Natur.493...62I}
{Ibata}, R.~A., {Lewis}, G.~F., {Conn}, A.~R., {Irwin}, M.~J., {McConnachie},
  A.~W., {Chapman}, S.~C., {Collins}, M.~L., {Fardal}, M., {Ferguson},
  A.~M.~N., {Ibata}, N.~G., {Mackey}, A.~D., {Martin}, N.~F., {Navarro}, J.,
  {Rich}, R.~M., {Valls-Gabaud}, D., \& {Widrow}, L.~M. 2013, \nat, 493, 62

\bibitem[{{Jerjen} {et~al.}(2000){Jerjen}, {Freeman}, \&
  {Binggeli}}]{2000AJ....119..166J}
{Jerjen}, H., {Freeman}, K.~C., \& {Binggeli}, B. 2000, \aj, 119, 166

\bibitem[{{Karachentsev}(2005)}]{2005AJ....129..178K}
{Karachentsev}, I.~D. 2005, \aj, 129, 178

\bibitem[{{Karachentsev} {et~al.}(2002){Karachentsev}, {Sharina}, {Dolphin},
  {Grebel}, {Geisler}, {Guhathakurta}, {Hodge}, {Karachentseva}, {Sarajedini},
  \& {Seitzer}}]{2002A&A...385...21K}
{Karachentsev}, I.~D., {Sharina}, M.~E., {Dolphin}, A.~E., {Grebel}, E.~K.,
  {Geisler}, D., {Guhathakurta}, P., {Hodge}, P.~W., {Karachentseva}, V.~E.,
  {Sarajedini}, A., \& {Seitzer}, P. 2002, \aap, 385, 21

\bibitem[{{Karachentseva} \& {Karachentsev}(1998)}]{1998A&AS..127..409K}
{Karachentseva}, V.~E. \& {Karachentsev}, I.~D. 1998, \aaps, 127, 409

\bibitem[{{Karachentseva} \& {Karachentsev}(2000)}]{2000A&AS..146..359K}
---. 2000, \aaps, 146, 359

\bibitem[{{Shaya} \& {Tully}(2013)}]{2013MNRAS.436.2096S}
{Shaya}, E.~J. \& {Tully}, R.~B. 2013, \mnras, 436, 2096

\bibitem[{{Tully}(2015)}]{2015AJ....149...54T}
{Tully}, R.~B. 2015, \aj, 149, 54

\bibitem[{{Tully} {et~al.}(2006){Tully}, {Rizzi}, {Dolphin}, {Karachentsev},
  {Karachentseva}, {Makarov}, {Makarova}, {Sakai}, \&
  {Shaya}}]{2006AJ....132..729T}
{Tully}, R.~B., {Rizzi}, L., {Dolphin}, A.~E., {Karachentsev}, I.~D.,
  {Karachentseva}, V.~E., {Makarov}, D.~I., {Makarova}, L., {Sakai}, S., \&
  {Shaya}, E.~J. 2006, \aj, 132, 729

\bibitem[{{Tully} {et~al.}(2008){Tully}, {Shaya}, {Karachentsev}, {Courtois},
  {Kocevski}, {Rizzi}, \& {Peel}}]{2008ApJ...676..184T}
{Tully}, R.~B., {Shaya}, E.~J., {Karachentsev}, I.~D., {Courtois}, H.~M.,
  {Kocevski}, D.~D., {Rizzi}, L., \& {Peel}, A. 2008, \apj, 676, 184

\end{thebibliography}

\bibliographystyle{apj}

\
\end{document}